\documentclass[aip, reprint]{revtex4-2}
\usepackage{amsfonts, amsmath, bm, amssymb, color}
\usepackage{graphicx}
\usepackage[colorlinks=true, citecolor=blue, linkcolor=blue]{hyperref}
\usepackage{booktabs}

\usepackage[T2A]{fontenc}
\usepackage[utf8]{inputenc}

\begin{document}

\title{Wave-driven geostrophic vortices in rotating turbulence}

\author{Vladimir Parfenyev}\email{parfenius@gmail.com}
\author{Sergey Vergeles}
\affiliation{Landau Institute for Theoretical Physics, Russian Academy of Sciences, 1-A Akademika Semenova av., 142432 Chernogolovka, Russia}
\affiliation{HSE University, Myasnitskaya 20, 101000 Moscow, Russia}

\date{\today}

\begin{abstract}
We perform direct numerical simulations of rotating turbulence to investigate the mechanisms responsible for the maintenance of large-scale geostrophic vortices. We confirm that the Reynolds stresses generated by inertial waves are consistent with the Reynolds-Averaged Navier-Stokes balance of the large-scale flow and demonstrate the predicted cyclone-anticyclone asymmetry, with cyclones exhibiting stronger azimuthal velocities and more compact cores. Wave-field analysis provides evidence that geostrophic vortices selectively reflect and absorb inertial waves according to the sign of their angular momentum. Wave absorption occurs either through the formation of a critical layer ahead of the reflection point or through enhanced viscous dissipation associated with increasing local wavenumber. As rotation increases, fewer inertial waves undergo resonant absorption, reducing angular-momentum transfer to the vortices and weakening the geostrophic flow. Consequently, the fraction of energy contained in geostrophic modes depends non-monotonically on the rotation rate, demonstrating that sufficiently rapid rotation can ultimately suppress, rather than enhance, turbulence two-dimensionalization.

\end{abstract}

\maketitle

\section{Introduction}

Understanding turbulence in rotating fluids is a longstanding challenge in hydrodynamics, driven by a wide range of geophysical, astrophysical, and engineering applications~\cite{davidson2013turbulence, greenspan1969theory, vallis2017atmospheric}. Rotation fundamentally alters the properties of turbulent flows and breaks the isotropy characteristic of classical three-dimensional turbulence when the Coriolis force dominates over inertial forces. As a result, rapidly rotating flows spontaneously develop large-scale quasi-two-dimensional vortical structures aligned with the rotation axis, as observed in laboratory experiments~\cite{mcewan1976angular, hopfinger1982turbulence, ruppert2005extraction, staplehurst2008structure, moisy2011decay, gallet2014scale, boffetta2020cyclone, tumachev2023two, tumachev2024observation} and numerical simulations~\cite{bartello1994coherent, yeung1998numerical, smith1999transfer, yoshimatsu2011columnar, biferale2016coherent, seshasayanan2018condensates}. At the same time, the small-scale dynamics are dominated by inertial waves, whose dispersion and interactions determine the transfer of energy across scales~\cite{galtier2003weak, yarom2014experimental, campagne2015disentangling, monsalve2020quantitative, brunet2020shortcut, shaltiel2024direct, shaltiel2025confirming}. Elucidating how these vortical and wave components coexist and exchange energy and momentum is one of the fundamental problems in the theory of rotating turbulence.

Particular attention has recently been devoted to the mechanisms that sustain coherent geostrophic vortices against viscous dissipation and Ekman friction. Since these structures contain a substantial fraction of the total kinetic energy, their persistence requires a continuous transfer of momentum from the surrounding turbulent fluctuations. A theoretical description of this process has emerged from quasi-linear theory, which assumes that nonlinear interactions among inertial waves are weak compared with their interactions with the intense large-scale vortices. Under this assumption, the wave dynamics can be treated within a linear framework, that substantially simplifies calculation of the Reynolds stress tensor generated by the waves. The resulting Reynolds stresses provide the feedback necessary to obtain a self-consistent equation governing the evolution and structure of the coherent geostrophic flow.

Early applications of this approach considered spatially homogeneous small-scale forcing and predicted the radial velocity profiles of geostrophic vortices in the asymptotic limit of vanishing Rossby number, where cyclones and anticyclones differ only by the sign of their rotation~\cite{kolokolov2020structure, parfenyev2021influence}. Extensions to finite Rossby numbers revealed a systematic cyclone-anticyclone asymmetry, predicting stronger and more compact cyclones than anticyclones~\cite{parfenyev2021velocity}. More recently, the interaction between inertial waves and geostrophic vortices has been investigated using a WKB description of wave propagation in slowly varying vortex flows~\cite{ivchenko2025absorption}. This analysis demonstrated that vortices interact with inertial waves through critical-layer absorption: inertial waves carrying angular momentum of the same sign as the vortex rotation are absorbed, whereas waves carrying angular momentum of the opposite sign are reflected from turning points. Consequently, geostrophic vortices can maintain themselves by selectively extracting angular momentum from the ambient wave field generated by the forcing. Recently, this theory was extended to an ensemble of incident waves and showed good agreement with experimental data~\cite{ivchenko2026maintenance}. A closely related mechanism has also been identified for coherent jet structures that replace geostrophic vortices in rotating periodic rectangular domains with unequal horizontal dimensions~\cite{gome2026waves}. Similar wave-mean-flow interactions are known to play a central role in stratified turbulence through the absorption of internal gravity waves at critical layers~\cite{booker1967critical}, highlighting the broader significance of these processes.

Despite these theoretical advances, direct numerical evidence supporting the proposed mechanisms remains limited. The primary objective of the present work is therefore to test and validate recent theoretical predictions using direct numerical simulations (DNS) of rotating turbulence. First, we measure the off-diagonal component of the Reynolds stress tensor and the velocity profiles of geostrophic vortices, demonstrating their consistency with the Reynolds-Averaged Navier-Stokes equation governing the large-scale quasi-two-dimensional flow~\cite{kolokolov2020structure, parfenyev2021influence, parfenyev2021velocity}. We then investigate the cyclone-anticyclone asymmetry predicted at finite Rossby numbers and confirm that cyclones exhibit larger peak azimuthal velocities, with the velocity maximum located closer to the vortex center than in anticyclones~\cite{parfenyev2021velocity}. Finally, we examine the propagation of inertial waves inside geostrophic vortices. Our observations show that the selective absorption of converging cylindrical waves carrying angular momentum of the same sign as the vortex rotation can occur through two distinct mechanisms. The first is absorption at a critical layer~\cite{ivchenko2025absorption}, while the second is a subcritical mechanism arising from enhanced viscous dissipation of the wave as the local wavenumber increases. A similar subcritical mechanism of wave-driven mean-flow generation has been proposed to explain quasi-biennial oscillations in stratified fluids~\cite{semin2024quasi}.

The selective nature of wave propagation has an unexpected consequence. Absorption occurs within critical layers where the Doppler-shifted frequency of an inertial wave vanishes. As the background rotation rate increases, this resonance condition becomes increasingly restrictive and can be satisfied by a progressively smaller subset of inertial waves. The selective enhancement of viscous dissipation is also reduced as the dispersion relations of waves carrying opposite signs of angular momentum become closer to each other. Consequently, the transfer of angular momentum from waves to vortices weakens, reducing the efficiency with which geostrophic structures are maintained. Thus, we predict a counterintuitive regime in which increasing the rotation rate beyond a certain threshold suppresses rather than enhances the two-dimensionalization of the flow. Our DNS results confirm this prediction and reveal a reduction in the fraction of flow energy contained in geostrophic modes at sufficiently rapid rotation.

\section{DNS Setup}

The numerical simulations are performed in a cubic domain of side length $L=2\pi$, rotating about the vertical $Z$-axis with angular velocity $\Omega$. Periodic boundary conditions are imposed in the horizontal $(X,Y)$ directions, thereby eliminating the formation of thin viscous boundary layers along vertical sidewalls. In contrast, periodic boundary conditions in the $Z$-direction would permit the development of vertical jet-like flows. Such structures are absent in laboratory experiments owing to the presence of impermeable top and bottom boundaries, but are observed in DNS when periodic boundary conditions are imposed in the vertical direction~\cite{biferale2016coherent}. Moreover, these vertical flows may interact with and absorb inertial waves, providing an additional pathway for wave momentum transfer and thereby complicating a direct comparison between theory and numerical simulations. To suppress the emergence of such large-scale vertical flows while simultaneously avoiding the strong velocity gradients associated with no-slip boundary layers at the top and bottom boundaries, we impose stress-free boundary conditions in the $Z$-direction.

To arrest the upscale energy transfer, we add a linear damping term $-\alpha \bm v$ to the Navier-Stokes equation, where $\bm v$ is the velocity field. In laboratory experiments with no-slip top and bottom boundaries, the upscale energy transfer is typically limited by Ekman friction $\alpha = 2\sqrt{\nu\Omega}/H$ before the energy-containing structures can grow to the system size, where $\nu$ is the kinematic viscosity and $H$ is the fluid depth~\cite{pedlosky2013geophysical,parfenyev2021influence}. As a result, geostrophic vortices remain smaller than the system size, with their number increasing and their size decreasing as $\Omega$ grows~\cite{tumachev2023two}. In DNS, we choose a relatively small friction coefficient $\alpha$, allowing the inverse cascade to reach the domain scale and produce system-size geostrophic vortices. This regime is particularly well suited for investigating wave-vortex interactions and testing the theoretical predictions addressed in this work.

Thus, the flow dynamics are governed by the Navier-Stokes equation in a rotating reference frame
\begin{equation}\label{eq:NS}
\partial_t \bm v + (\bm v \cdot \nabla) \bm v + 2 [\bm \Omega \times \bm v] = - \nabla p + \nu \nabla^2 \bm v - \alpha \bm v + \bm f
\end{equation}
subject to the incompressibility condition $\nabla \cdot \bm v = 0$. Here, the kinematic pressure $p$ includes both the gravitational and centrifugal potentials. We employ a zero-mean, Gaussian, white-in-time forcing $\bm f$, generated independently at each numerical time step. The forcing is confined to Fourier modes in the wavenumber range $k_f \in [5,7]$, projected onto its divergence-free component, and subsequently transformed to physical space. Its amplitude is normalized to yield a prescribed energy-injection rate $\varepsilon = \langle \bm v \cdot \bm f \rangle_{S,T}$, where $\langle \cdot \rangle_{S,T}$ denotes averaging over physical space and time. Unless otherwise stated, the simulations are performed with rotation rate $\Omega=8$, linear friction coefficient $\alpha = 5 \times 10^{-3}$, kinematic viscosity $\nu = 2 \times 10^{-3}$, and forcing power $\varepsilon = 0.1$.

The DNS are carried out using the software package Oceananigans.jl~\cite{wagner2025high}. The code solves Eq.~(\ref{eq:NS}) on GPU using a finite-volume discretization. Advection terms are approximated with a centered second-order scheme, while time integration is performed using a third-order Runge-Kutta method. The integration time step $dt$ is adjusted adaptively to maintain a Courant number $\lesssim 0.5$, ensuring numerical stability and convergence. We additionally verified that $dt$ remains much smaller than both the inertial-wave period $(2 \Omega)^{-1}$ and the Kolmogorov time scale $\tau_{\eta} = (\nu/\varepsilon)^{1/2}$. All simulations employ a uniform Cartesian grid with a resolution of $N^3 = 512^3$. The grid spacing $dx=L/N$ is sufficiently fine to resolve the Kolmogorov dissipation scale $\eta = (\nu^3/\varepsilon)^{1/4}$; in our simulations $\eta/dx \approx 1.37$ and we also report the grid-convergence tests in the Supplemental Material (SM)~\cite{SM}. To reduce memory consumption on GPU, all model fields are stored in single-precision format.

In the following analysis, the Cartesian components of the velocity field $\bm v$ are denoted by $(u,v,w)$. We further decompose $\bm v$ into a vertically averaged component and a residual three-dimensional component
\begin{equation}\label{eq:decomposition}
\bm v = \bar{\bm v} + \bm v_{3D},
\end{equation}
where the averaging along the rotation axis is defined as $\bar{(\cdot)} = (1/L) \int_0^{L} dz (\cdot)$. This decomposition provides a convenient framework for distinguishing between vertically averaged and three-dimensional motions and for quantifying their respective contributions to the flow dynamics.

\section{Self-Organization into Geostrophic Vortices}

First, we impose the zero initial condition $\bm v = 0$ and integrate the system up to $T=400$, which is sufficient for the flow to reach a statistically stationary state. Snapshots of the velocity field are saved every $\Delta t = 20$ time units throughout this initial transient stage.

\begin{figure}[b]
    \centering
    \includegraphics[width=0.9\linewidth]{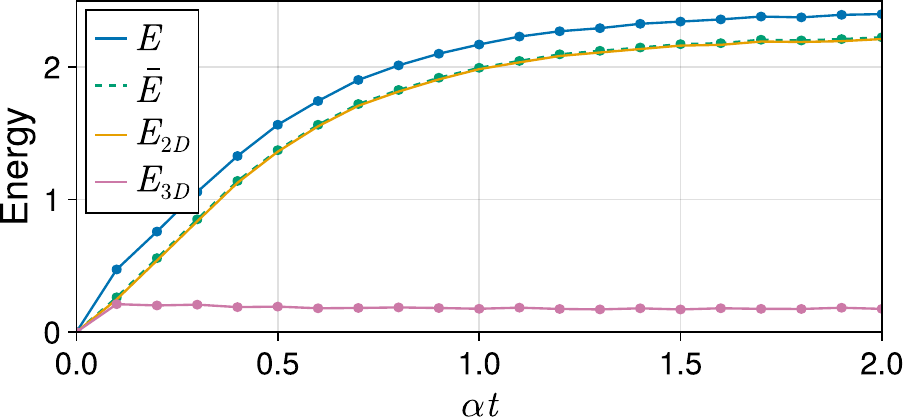}
    \caption{Kinetic energy of the system as a function of time.}
    \label{fig:1}
\end{figure}

Figure~\ref{fig:1} shows the time evolution of the kinetic energy $E = \langle u^2 + v^2 + w^2\rangle_S/2$. The figure also displays the contributions associated with vertically invariant modes, horizontal geostrophic currents only, and the residual three-dimensional motions
\begin{equation}
\begin{gathered}
\bar{E} = \dfrac{\langle \bar u^2 + \bar v^2 + \bar w^2\rangle_S}{2},
\quad
E_{2D} = \dfrac{\langle \bar u^2 + \bar v^2\rangle_S}{2},
\\
E_{3D} = \dfrac{\langle u_{3D}^2 + v_{3D}^2 + w_{3D}^2\rangle_S}{2},
\end{gathered}
\end{equation}
where $\langle \cdot \rangle_S = (1/L^3) \int d^3 \bm r \,(\cdot)$ denotes the space averaging. It follows directly from the decomposition (\ref{eq:decomposition}) that $E = \bar{E} + E_{3D}$. As seen in Fig.~\ref{fig:1}, the flow reaches a statistically stationary state on a timescale of order $t \sim 1/\alpha$. In this regime, 92.7\% of the total kinetic energy is contained in the vertically invariant modes, while the remaining 7.3\% is associated with three-dimensional motions. Owing to the stress-free boundary conditions in the $Z$-direction, vertical jet-like motions are strongly suppressed. Their contribution is small both relative to the horizontal geostrophic flow, $2 \langle \bar w^2 \rangle_S/ \langle \bar u^2 + \bar v^2 \rangle_S \approx 0.01$, and relative to the three-dimensional component, $3 \langle \bar w^2 \rangle_S/ \langle u_{3D}^2 + v_{3D}^2 + w_{3D}^2 \rangle_S \approx 0.2$.

Next, using the root-mean-square velocity $U_{rms} = \sqrt{2E}$ and the forcing length scale $l_f = 2 \pi /k_f$, we estimate the Reynolds and Rossby numbers of the statistically stationary flow:
\begin{equation}
Re = \dfrac{U_{rms} l_f}{\nu} \approx 1150, \quad Ro = \dfrac{U_{rms}}{2 \Omega l_f} \approx 0.13.
\end{equation}
The large Reynolds number indicates a turbulent flow, while the small Rossby number implies that rotation strongly affects motions at the forcing scale and larger. According to the Taylor-Proudman theorem~\cite{proudman1916motion, taylor1917motion}, this suppresses variations along the rotation axis and promotes two-dimensionalization of the flow. Under these conditions, one expects an upscale transfer of kinetic energy toward larger scales. Furthermore, the large value of $Rh = U_{rms}/(\alpha L) \approx 70$ indicates weak large-scale friction, allowing the formation of system-size quasi-two-dimensional geostrophic vortices.

\begin{figure}[t]
    \centering
    \includegraphics[width=0.8\linewidth]{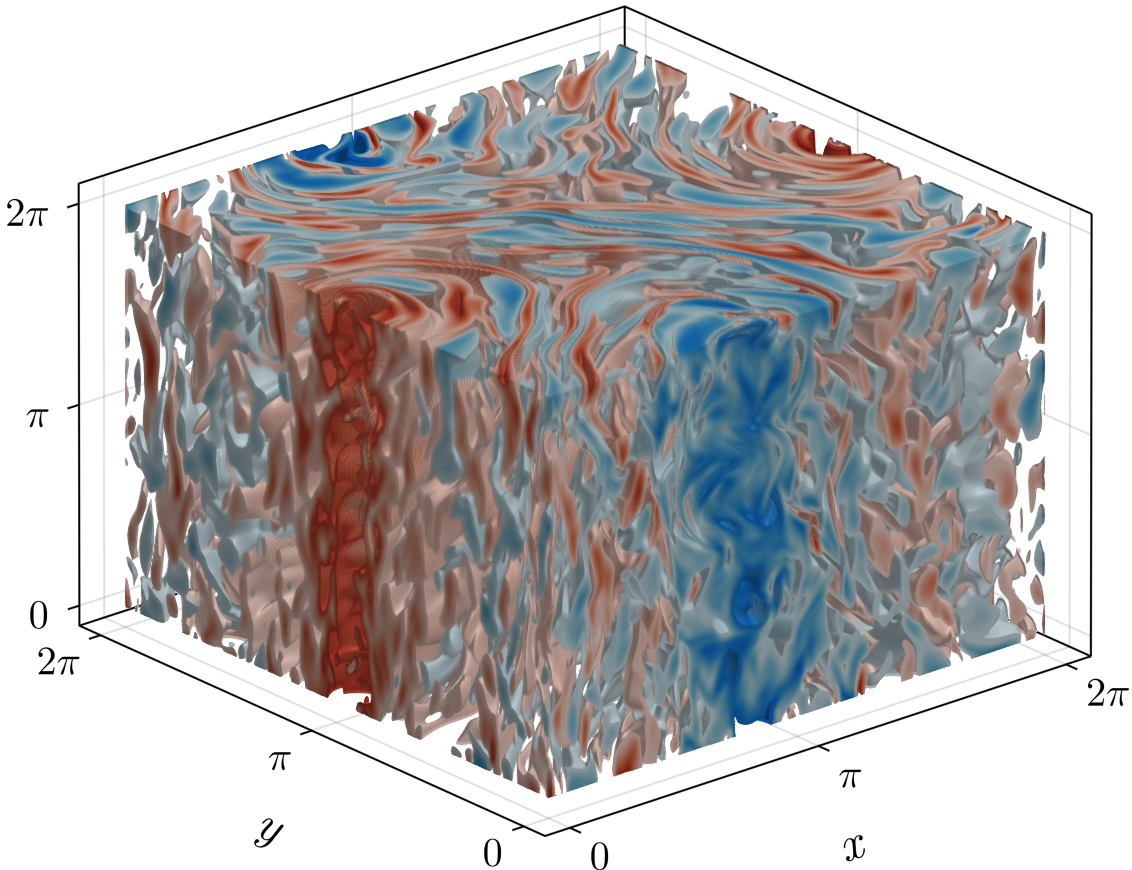}
    \caption{Distribution of vertical vorticity. Red (blue) denotes cyclonic (anticyclonic) rotation.}
    \label{fig:2}
\end{figure}

Figure~\ref{fig:2} shows the distribution of vertical vorticity $\zeta = \partial_x v - \partial_y u$ in the statistically stationary state. Two large-scale columnar geostrophic vortices are clearly visible on the faces of the computational domain: a cyclone (red) and an anticyclone (blue). Their size is comparable to that of the domain, consistent with the development of an inverse energy transfer and the accumulation of kinetic energy at the largest available scales. The vortices are not stationary but undergo a slow drift in the horizontal plane while largely preserving their structure. Note that similar wandering motions have been reported in studies of purely two-dimensional turbulence, where their statistical properties were investigated~\cite{parfenyev2024statistical, chan2012dynamics}.

\section{Energy Spectra and Fluxes}

We now turn to the description of the system in its statistically stationary state. To collect sufficient statistics, we performed eight independent simulations, each of duration $T=1000$. Velocity fields were stored every $\Delta t = 2$, which is approximately equal to the turnover time of a geostrophic vortex. For reference, the integration time step $dt=8 \times 10^{-4}$. Different realizations of the stochastic forcing were used in each simulation, while the initial condition was taken as the final velocity-field snapshot from the previous section, see Fig.~\ref{fig:2}. The resulting dataset is more than sufficient for averaging sign-definite quantities such as energy spectra or radial velocity profiles of geostrophic vortices. However, this large statistics will be needed later to analyze the cross-component of the Reynolds stress tensor generated by turbulent fluctuations, which is known to exhibit strong variability~\cite{frishman2018turbulence}.

\begin{figure}[t]
    \centering
    \includegraphics[width=0.9\linewidth]{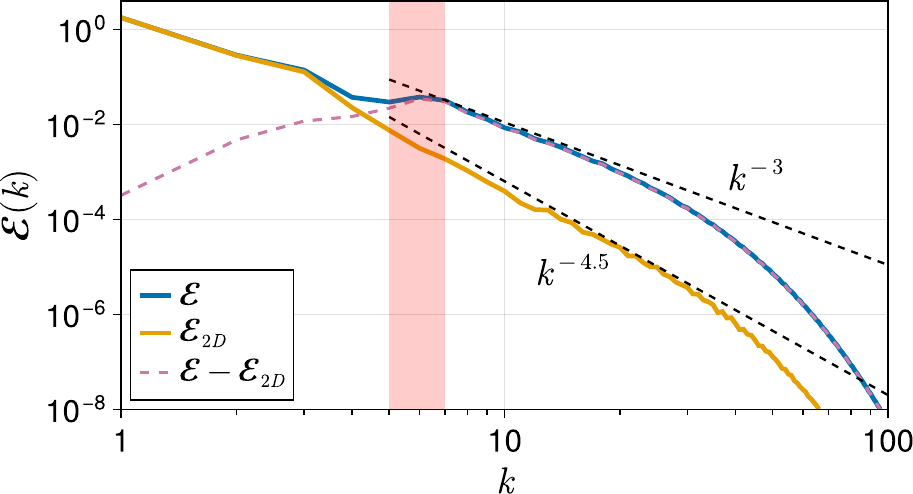}
    \caption{Energy spectra of the total velocity field $\mathcal{E}(k)$, the geostrophic component $\mathcal{E}_{2D}(k)$, and their difference.}
    \label{fig:3}
\end{figure}

To compute the kinetic energy spectrum, the velocity field $\bm v=(u,v,w)$ is decomposed into basis functions that satisfy the boundary conditions. Owing to periodicity in the horizontal directions, the velocity field is represented by Fourier modes $e^{i k_x x+ i k_y y}$. In the vertical direction, the stress-free boundary conditions require the horizontal velocity components $u$ and $v$ to be expanded in cosine modes, while the vertical velocity component $w$ is expanded in sine modes:
\begin{eqnarray}
    &\displaystyle u(x,y,z) = \sum_{k_x,k_y} \sum_{k_z \geq 0} \hat u(\bm k) e^{i k_x x + i k_y y} \cos(k_z z),&\\
    &\displaystyle v(x,y,z) = \sum_{k_x,k_y} \sum_{k_z \geq 0} \hat v(\bm k) e^{i k_x x + i k_y y} \cos(k_z z),&\\
    &\displaystyle w(x,y,z) = \sum_{k_x,k_y} \sum_{k_z > 0} \hat w(\bm k) e^{i k_x x + i k_y y} \sin(k_z z),&
\end{eqnarray}
where $\bm k = (k_x, k_y, k_z)$ denotes the wave vector, $k_x = 2 \pi n_x/L$, $k_y = 2 \pi n_y/L$, $k_z = \pi n_z/L$, and $n_x, n_y \in \mathbb{Z}$, while $n_z \in \mathbb{Z}_{\geq 0}$. The distribution of kinetic energy across spatial scales is then characterized by the energy spectrum (averaged over spherical shells, despite its anisotropy associated with the direction of rotation)
\begin{equation}
\mathcal{E}(k) = \sum_{k-1 < |\bm k| \leq k} W(k_z)|\hat{\bm v}(\bm k)|^2/2,
\end{equation}
where $\hat{\bm v} = (\hat u, \hat v, \hat w)$, and the weight factor $W(k_z)$ accounts for the normalization of the trigonometric basis in the vertical direction, with $W(0) = 1$ and $W(k_z) = 1/2$ for all $k_z > 0$. The contribution of the horizontal geostrophic flow is quantified separately through the two-dimensional energy spectrum
\begin{equation}
\mathcal{E}_{2D}(k_h) = \sum_{k_h-1 < |\bm k_h| \leq k_h} \left( |\hat{\bar u}(\bm k_h)|^2 + |\hat{\bar v}(\bm k_h)|^2 \right)/2,
\end{equation}
where $\bm k_h = (k_x, k_y)$ is the horizontal wave vector, $\bar u(x,y)$ and $\bar v(x,y)$ are the vertically averaged horizontal velocity components, and the corresponding spectral coefficients $\hat{\bar u}(\bm k_h)$ and $\hat{\bar v}(\bm k_h)$ are obtained via a standard 2D Fast Fourier Transform.

The results are presented in Fig.~\ref{fig:3}. They show that the dominant contribution to the total energy comes from the large scales, which are associated with the quasi-two-dimensional geostrophic modes, in agreement with the results reported in Fig.~\ref{fig:1}. The spectral shape in this range is apparently determined by the mean velocity profiles of the geostrophic vortices, similarly to fully two-dimensional turbulent flows~\cite{parfenyev2024statistical, van2025two}. The scales smaller than the forcing scale $l_f = 2 \pi/k_f$ are strongly three-dimensional. For $k>k_f$, we observe dependencies close to $\mathcal{E}(k) \propto k^{-3}$ and $\mathcal{E}_{2D}(k) \propto k^{-4.5}$. We note that a similar scaling, $\mathcal{E}(k) \propto k^{-3}$, was previously reported in Ref.~\onlinecite{seshasayanan2018condensates} for a system with periodic boundary conditions in all directions and without additional linear drag. In the SM~\cite{SM}, we further discuss how the energy spectra are modified when a standard Fast Fourier Transform is formally applied in the $Z$-direction despite the lack of periodicity. We also examine the effect of applying a Hann window, commonly used in experimental data processing, on the resulting spectra in this case.

\begin{figure}[t]
    \centering
    \includegraphics[width=0.85\linewidth]{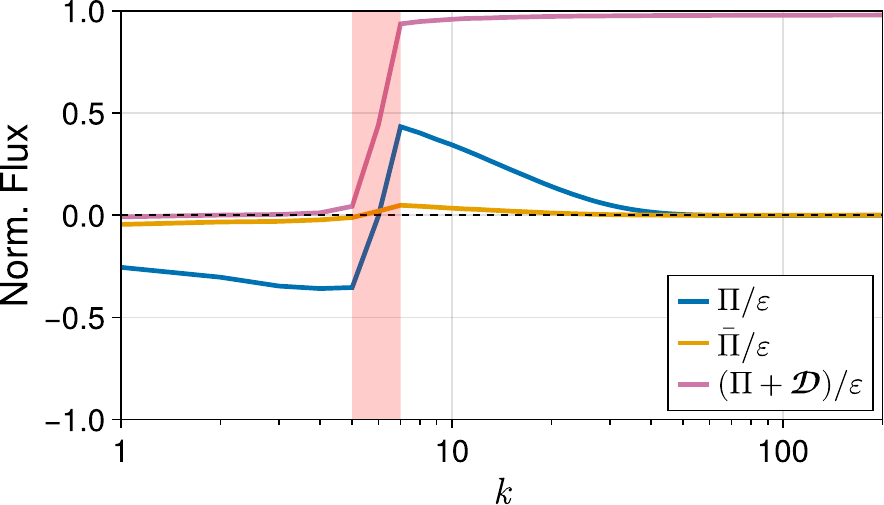}
    \caption{Normalized energy fluxes associated with the total velocity field $\Pi(k)/\varepsilon$ and with the vertically averaged modes $\bar{\Pi}(k)/\varepsilon$ shown alongside with the normalized cumulative power $\mathcal{F}(k)/\varepsilon=\Pi(k)/\varepsilon + \mathcal{D}(k)/\varepsilon$ injected into modes with wavenumbers $\leq k$.}
    \label{fig:4}
\end{figure}

Next, we examine the energy fluxes in the system associated with the full velocity field and with the vertically averaged modes alone~\cite{seshasayanan2018condensates}:
\begin{equation}
\begin{gathered}
    \Pi(k) = \langle v_i^{k<} v_j \partial_j v_i \rangle_{S,T},\\
    \bar{\Pi} (k_h) = \langle \bar v_i^{k_h<} \bar v_{\beta} \partial_{\beta} \bar v_i \rangle_{S,T}.
\end{gathered}
\end{equation}
Here, $\bm v^{k<}$ and $\bar{\bm v}^{k_h<}$ denote the filtered velocity fields (in physical space) containing only Fourier modes with wavenumbers $\leq k$ and $\leq k_h$, correspondingly. Latin indices run over the three spatial coordinates $\{x,y,z\}$, while Greek indices take values $\{x,y\}$. Repeated indices imply summation according to the Einstein convention. With this definition, the positive/negative fluxes correspond to downscale/upscale energy transfer.

The resulting energy fluxes are shown in Fig.~\ref{fig:4}. As expected, for $k>k_f$ the energy undergoes a forward cascade toward smaller scales, whereas for $k<k_f$ the flux is directed toward larger scales, indicating an inverse (upscale) energy transfer. Notably, both energy fluxes are predominantly mediated by the three-dimensional motions, while the contribution of the vertically averaged modes, represented by $\bar{\Pi}$, remains comparatively small. In addition, we compute the cumulative dissipation,
\begin{equation}
\mathcal{D}(k) = \sum_{|\bm q|\leq k} 2(\alpha + \nu q^2) \mathcal{E} (\bm q),
\end{equation}
which accounts for energy loss due to linear friction and viscous dissipation. In the statistically stationary state, the energy budget satisfies $\Pi(k) + \mathcal{D}(k) = \mathcal{F}(k)$,
where $\mathcal{F}(k)$ is the cumulative power injected into modes with $|\bm q| \leq k$. As expected, $\mathcal{F}(k)$ saturates at the energy injection rate $\varepsilon$ for $k>k_f$, demonstrating that the numerical simulation satisfies the global energy balance.

\section{Structure of Geostrophic Vortices}

\begin{figure}[b]
    \centering
    \includegraphics[width=0.8\linewidth]{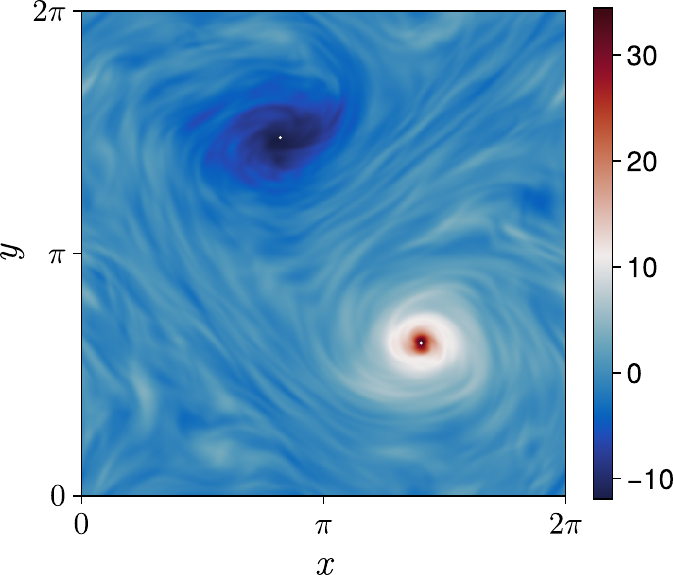}
    \caption{The white dots mark the axes of the geostrophic vortices. Red (blue) denotes cyclonic (anticyclonic) rotation.}
    \label{fig:5}
\end{figure}

In this section, we analyze the structure of coherent columnar vortices. Our primary focus is the velocity profiles within cyclone and anticyclone. To measure them, we adopt reference frames moving with the corresponding columnar vortices. We first compute the vertical vorticity $\zeta = \partial_x v - \partial_y u$ and then average it over the $Z$-direction. To suppress small-scale turbulent fluctuations, the resulting field $\bar \zeta (x,y)$ is smoothed using a Gaussian filter with $\sigma = 15$ pixels for cyclone and $\sigma=30$ pixels for anticyclone. The locations of the geostrophic vortex axes are then identified as the maximum and minimum of the filtered vorticity field. An example of the vortex axis detection procedure is shown in Fig.~\ref{fig:5}, where the color map represents the $Z$-averaged vertical vorticity $\bar \zeta(x,y)$ before the smoothing filter is applied. A clear asymmetry between cyclones and anticyclones is evident: the cyclone is more compact and exhibits higher vorticity at its core.

\begin{figure*}[t]
    \centering
    \includegraphics[width=0.85\linewidth]{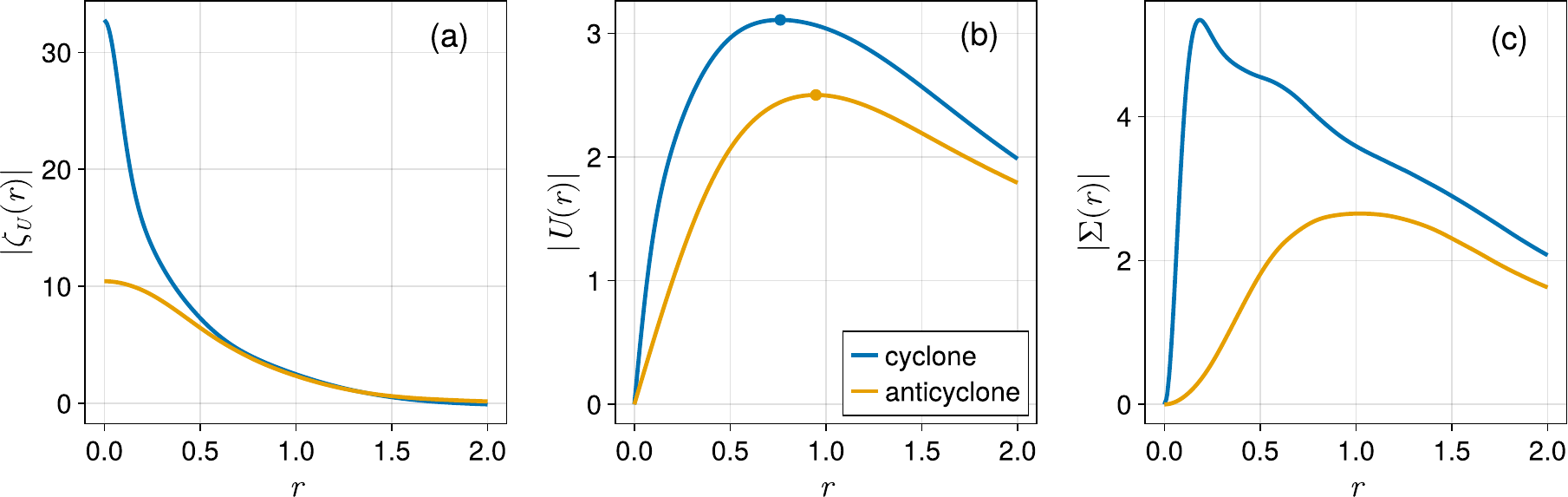}
    \caption{Radial profiles of (a) vorticity, (b) azimuthal velocity, and (с) local shear rate for coherent geostrophic vortices.}
    \label{fig:6}
\end{figure*}

Next, we identify the vortex axes in each saved snapshot of $\bar \zeta(x,y)$ and separately align the centers of cyclones and anticyclones. Averaging over all snapshots (i.e., over time) yields the mean vorticity profiles, which are axisymmetric and denoted by $\zeta_U(r)$. According to Stokes' theorem, the corresponding mean azimuthal velocities are given by
\begin{equation}
U(r) = \int_0^r dx \, \zeta_U(x) x/r. 
\end{equation}
The rotation within geostrophic vortices is differential, and one of its key characteristics is the local shear rate
\begin{equation}\label{Sigma-def}
\Sigma (r) = r \partial_r (U/r) = \zeta_U - 2U/r, 
\end{equation}
which transfers energy from the three-dimensional fluctuations to the coherent vortex, thereby sustaining its existence~\cite{kolokolov2020structure, parfenyev2021influence, parfenyev2021velocity}. All of these quantities were measured in DNS, and the results are presented in Fig.~\ref{fig:6}. We observe that the azimuthal velocity is higher in cyclone than in anticyclone, and that its maximum is located closer to the vortex axis in cyclone. These findings support qualitatively the theoretical predictions of Ref.~\onlinecite{parfenyev2021velocity}. At the same time, the detailed vortex structure deviates from the asymptotic theory~\cite{parfenyev2021influence}. This discrepancy is likely due to the relatively large forcing scale in our simulations, which is comparable to the vortex size and therefore violates the scale separation assumed in the theory. A similar effect has previously been reported for purely two-dimensional coherent vortices~\cite{parfenyev2022profile}. 

Rather than comparing our results with the predictions of the asymptotic theory, we can directly analyze the equation governing the structure of geostrophic vortices. To derive this equation, we introduce a cylindrical coordinate system centered on the axis of a geostrophic vortex and decompose the velocity field as
\begin{equation}
\bm v= \bm u+\bm U,\quad \bm u=\{u_{{r}}, u_{\varphi}, u_z\},\quad \bm U=\{0, U, 0\},
\end{equation}
where $\bm U$ denotes the mean azimuthal flow and $\bm u$ represents the velocity fluctuations superimposed on it. Substituting this decomposition into the Navier-Stokes equation (\ref{eq:NS}), averaging over the fast turbulent fluctuations, and projecting onto the azimuthal direction, we obtain~\cite{parfenyev2021influence, parfenyev2021velocity}
\begin{equation}
\partial_t U = - \alpha U -\left( \partial_{{r}} + \frac{2}{{r}} \right) \left( \langle u^{\varphi} u^{{r}} \rangle_{Z,\Phi,T} - \nu \Sigma \right),
\end{equation}
where $\langle \cdot \rangle_{Z,\Phi,T}$ denotes averaging over the vertical direction, the polar angle, and time. In the statistically stationary state, the time derivative $\partial_t U$ can be neglected. Integrating the resulting equation then yields
\begin{equation}\label{eq:Reynolds}
\langle u^{\varphi} u^{{r}} \rangle_{Z,\Phi,T} - \nu \Sigma + \alpha \int_0^r dx \, U(x) x^2/r^2 = 0.
\end{equation}
The last two terms represent the dissipation associated with fluid viscosity and linear friction, respectively. This dissipation is balanced by the off-diagonal component of the Reynolds stress tensor $\langle u^{\varphi} u^{{r}} \rangle_{Z,\Phi,T}$, which continuously transfers momentum from the turbulent fluctuations to the mean flow and thereby sustains the coherent geostrophic vortices.

\begin{figure}[b]
    \centering
    \includegraphics[width=0.9\linewidth]{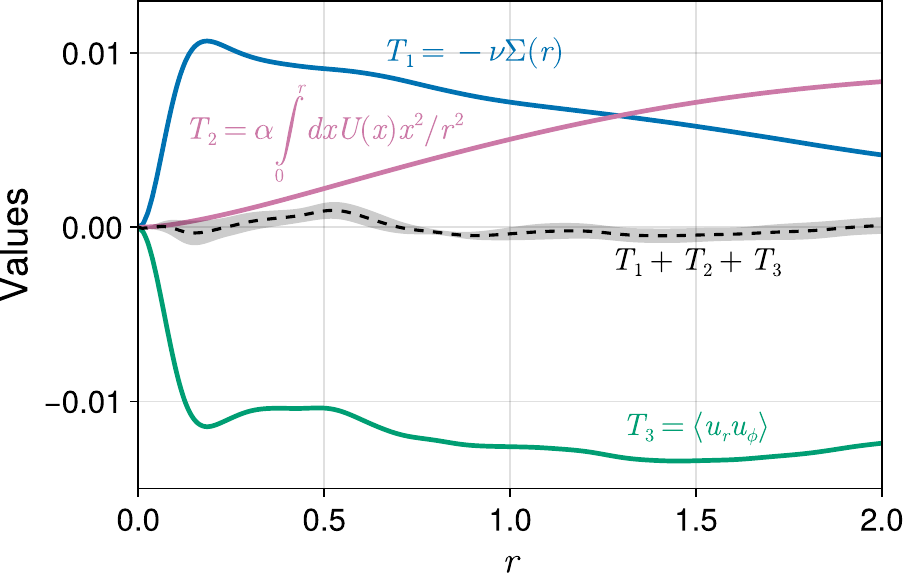}
    \caption{Validation of Eq.~(\ref{eq:Reynolds}) for the cyclone. Averages are computed over eight independent simulations ($4000$ snapshots in total). The shaded region denotes the standard deviation estimated from the eight runs using the central limit theorem.}
    \label{fig:7}
\end{figure}

To verify Eq.~(\ref{eq:Reynolds}), we measured each of its three terms directly from the DNS data. As noted above, the main challenge is the accurate determination of the Reynolds stress $\langle u^{\varphi} u^{{r}} \rangle_{Z,\Phi,T}$, because this quantity exhibits strong variability and therefore requires extensive statistical averaging to converge~\cite{frishman2018turbulence}. The results for the cyclone are presented in Fig.~\ref{fig:7} and show quantitative agreement with Eq.~(\ref{eq:Reynolds}). The corresponding analysis for the anticyclone yields similar results and is provided in the SM~\cite{SM}.

\section{Frequency Spectra}

\begin{figure}[b]
    \centering
    \includegraphics[width=0.9\linewidth]{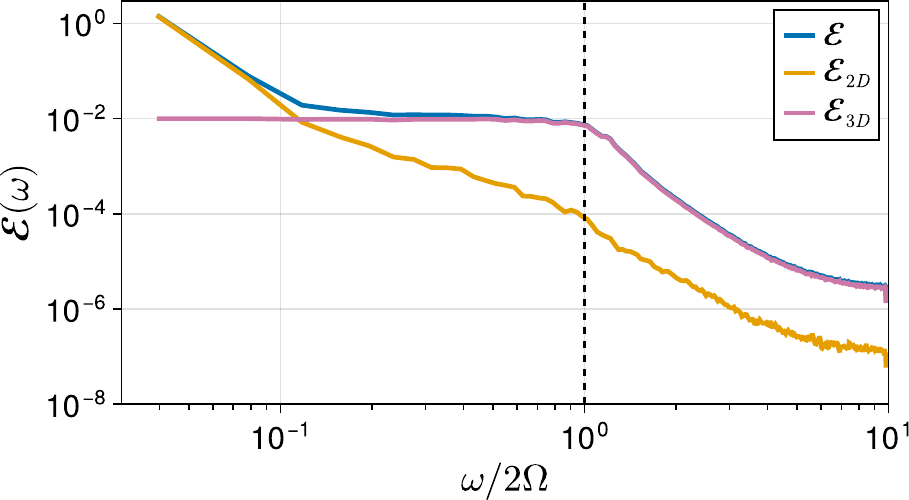}
    \caption{Frequency spectra of the total velocity field $\mathcal{E}(\omega)$, the geostrophic component $\mathcal{E}_{2D}(\omega)$, and the three-dimensional fluctuations $\mathcal{E}_{3D}(\omega)$. The vertical dashed line denotes the maximum frequency of inertial waves $\omega = 2\Omega$.}
    \label{fig:8}
\end{figure}

We now turn to the analysis of the frequency spectra of the system. Our main interest lies in the relatively fast three-dimensional motions. To resolve these dynamics, we perform an additional numerical simulation in the statistically stationary regime with a duration of $T=10$, during which the velocity field is recorded with a fine temporal resolution of $\Delta t = 0.02$. The frequency spectra of the total velocity field $\mathcal{E} (\omega)$, the geostrophic component $\mathcal{E}_{2D} (\omega)$, and the three-dimensional fluctuations $\mathcal{E}_{3D} (\omega)$ are shown in Fig.~\ref{fig:8}.

At low frequencies, the spectrum is dominated by the geostrophic contribution $\mathcal{E}_{2D}$. Superimposed on this background is the three-dimensional component $\mathcal{E}_{3D}$, which is associated with inertial waves. In the frequency range $\omega < 2 \Omega$, the spectral density of this component is approximately constant, reflecting the short temporal correlation of the forcing in our simulations. We do not observe signatures of weak inertial-wave turbulence~\cite{galtier2003weak, shaltiel2026evidence}, as the measured spectrum does not follow the theoretical prediction $\mathcal{E}_{3D} \propto \omega^{-1/2}$. This is most likely because the waves interact primarily with the relatively strong geostrophic flow, while wave-wave interactions remain of secondary importance.

For frequencies exceeding $\omega > 2 \Omega$, the three-dimensional component $\mathcal{E}_{3D}$ decreases sharply, consistent with the cutoff imposed by the inertial-wave dispersion relation $\omega = \pm 2\Omega k_z/k$. At these higher frequencies, the total spectrum is determined almost entirely by the three-dimensional modes, indicating that the geostrophic component evolves on much longer time scales and contributes negligibly to the high-frequency dynamics.

\section{Wave-vortex Interaction}

In this section, we investigate the selective absorption and reflection of inertial waves by geostrophic vortices, a mechanism described theoretically in Ref.~\onlinecite{ivchenko2025absorption}. To analyze wave propagation, we decompose the wave field into cylindrical harmonics of the form $\propto e^{im\varphi + ik_z z - i\omega t}$ in a reference frame whose vertical $Z$-axis is aligned with the axis of the geostrophic vortex. In the linear regime, an inertial wave propagating through the geostrophic flow is governed by the pressure-field wave equation (A6) of Ref.~\onlinecite{ivchenko2025absorption}. The corresponding vertical velocity component is then given by $w = k_z p/\widetilde\omega$. Here, $\widetilde{\omega} = \omega - mU/r$ is the intrinsic wave frequency, incorporating the Doppler shift induced by the geostrophic flow $U(r)$. This shift depends on the azimuthal wavenumber $m\in\mathbb{Z}$ of the cylindrical wave.

In order to qualitatively analyze the wave equation, we consider the short-wave limit. Then the equation is simplified to 
\begin{equation}\label{eq:m=0}
    \partial_r^2 p 
    +
    \frac{1}{r}\partial_rp
    +
    k_r^2p
    =
    0,
\end{equation}
with the dispersion relation
\begin{equation}\label{eq:dispersion}
    \widetilde \omega^2 
    = 
    \dfrac{2 \widetilde \Omega (2 \widetilde \Omega + \Sigma) k_z^2}{k^2},
    \quad 
    k^2 = k_r^2(r) + \frac{m^2}{r^2} + k_z^2.
\end{equation}
Here $\tilde \Omega = \Omega + U/r$ and $\Sigma$ is the local shear rate defined in Eq.~(\ref{Sigma-def}). Equation (\ref{eq:m=0}) describes wave oscillations in an axially symmetric potential. The term with the first derivative is held in order to reproduce the correct behavior near $r \to 0$. In particular, if the mean vortex flow is absent, equation (\ref{eq:m=0}) becomes exact and its solution is the Bessel function of the $m$-th order, $w \propto {\mathrm J}_m(\sqrt{k^2-k_z^2}\,r)$.

\begin{figure*}
    \noindent 
    \includegraphics[width=0.9\linewidth]{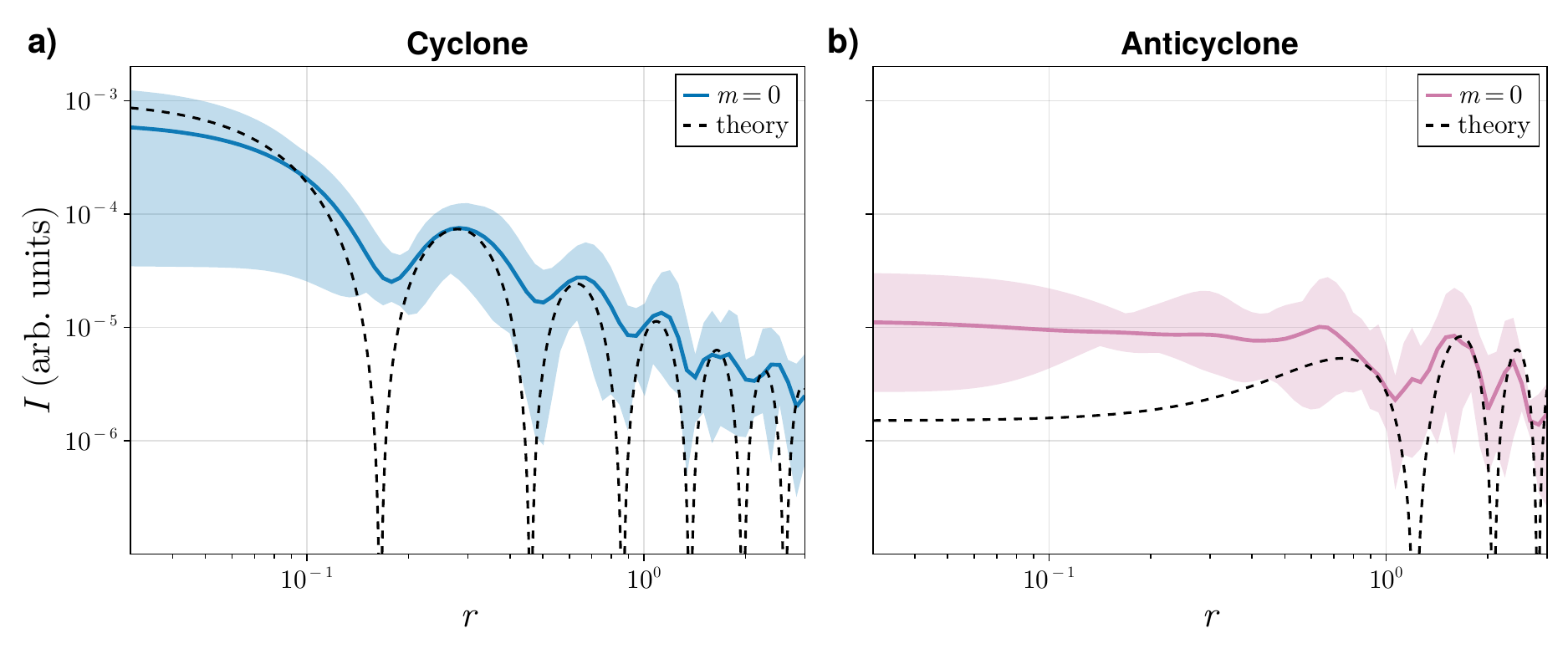}
    \caption{Radial profiles of the spectral intensity (\ref{eq:SI}) of the axisymmetric inertial-wave field ($m=0$, $k_z=4$, $\omega=10.6$) for a cyclone (a) and an anticyclone (b). Solid lines show DNS data with standard-deviation shading, while dashed lines show numerical solutions of the wave equation (A6) of Ref.~\onlinecite{ivchenko2025absorption}. Vertical lines indicate the critical layer or turning point, where applicable.}
    \label{fig:9}
\end{figure*}

The dispersion law (\ref{eq:dispersion}) exhibits two features that merit further discussion. The first concerns the formation of critical layers at a radial distance $r_*$ from the vortex axis, defined by the condition $\widetilde \omega (r_*) = 0$, or equivalently,
\begin{equation}\label{eq:crl}
    \omega - m\dfrac{U(r_*)}{r_*} = 0.
\end{equation}
According to the dispersion law, $k (r_*) \to \infty$, indicating that the wavenumber diverges at the critical layers in the inviscid limit. Finite fluid viscosity regularizes this singular behavior. Nevertheless, the qualitative picture remains important: the large wavenumber near a critical layer causes the viscous term in the dynamical equations to become dominant, resulting in the absorption of inertial waves and transfer their angular momentum to the geostrophic flow. Notably, this absorption is selective with respect to the sign of the azimuthal wavenumber $m$: only harmonics satisfying $\omega m U>0$ are absorbed. For example, inertial waves with positive frequency $\omega>0$ are absorbed inside cyclones ($U>0$) only when their azimuthal wavenumber $m>0$. Conversely, inside anticyclones ($U<0$) absorption occurs only for waves with $m<0$.

For weak geostrophic currents and harmonics with small azimuthal wavenumber $m$, the resonance condition (\ref{eq:crl}) may have no solution even for harmonics with the appropriate sign of $m$. In this case, according to the dispersion law (\ref{eq:dispersion}), the reduction of the intrinsic wave frequency $\widetilde{\omega}(r)$ is accompanied by a finite increase in $k(r)$, resulting in partial wave absorption. Note also, that near critical layers the inertial-wave wavelength decreases, naturally establishing a separation of scales between the geostrophic flow and the absorbed waves assumed in the theory.

The second feature of the dispersion law is associated with WKB turning points and the reflection of inertial waves from these locations. The positions $r_t$ of the turning points can be determined by imposing the condition $k_r(r_t)=0$ in the dispersion law (\ref{eq:dispersion}). Harmonics with either sign of $m$ possess WKB turning points, although their locations differ.

Unlike absorbed waves, reflected waves do not transfer their angular momentum to the geostrophic flow~\cite{ivchenko2025absorption}. In both cases, the wave amplitude is expected to decrease within the corresponding forbidden regions, $r<r_*$ for absorbed waves and $r<r_t$ for reflected waves. However, beyond the turning point $r>r_t$, reflected waves should exhibit the characteristic pattern of nodes and antinodes associated with standing waves. This feature can therefore be used to qualitatively distinguish between absorbed and reflected waves.

To characterize the propagation of inertial waves in DNS, we analyze the vertical velocity component $w (r, \varphi,z,t)$. To isolate individual wave modes, we compute the temporal Fourier transform using a Hann window, expand the vertical dependence in sine modes, and project the horizontal structure onto azimuthal Fourier harmonics:
\begin{equation}
w (r, \varphi,z,t)= \hspace{-0.5em} \sum_{m,k_z, \omega} \hspace{-0.2em} \hat w(r,m,k_z,\omega) e^{i m \varphi - i \omega t} \sin (k_z z).
\end{equation}
In our analysis, we focus on the harmonics with $k_z=4$ and frequencies close to $\omega = 2 \Omega k_z/k_f \approx 10.6$, consistent with the inertial-wave dispersion relation outside the geostrophic flow. To collect statistics, we perform eight independent simulations in the statistically stationary regime, each of duration $T=30$. During each run, the velocity field is recorded at intervals of $\Delta t=0.06$, corresponding to approximately ten samples per the wave period. The wave propagation is quantified in terms of the spectral intensity
\begin{equation}\label{eq:SI}
I(r,m)=\left\langle
\sum_{\sigma = \omega-\Delta\omega}^{\omega+\Delta\omega}
\left|\hat w(r,m,k_z,\sigma)\right|^2
\right\rangle_R,
\end{equation}
where the summation is performed over the frequency band $\Delta\omega=2\pi/T\approx0.21$ to improve statistical convergence, and the angle brackets $\langle \cdot \rangle_R$ denote averaging over the eight independent DNS runs. The the standard deviations reported below are estimated from these eight runs using the central limit theorem.

\begin{figure*}
    \noindent 
    \includegraphics[width=0.9\linewidth]{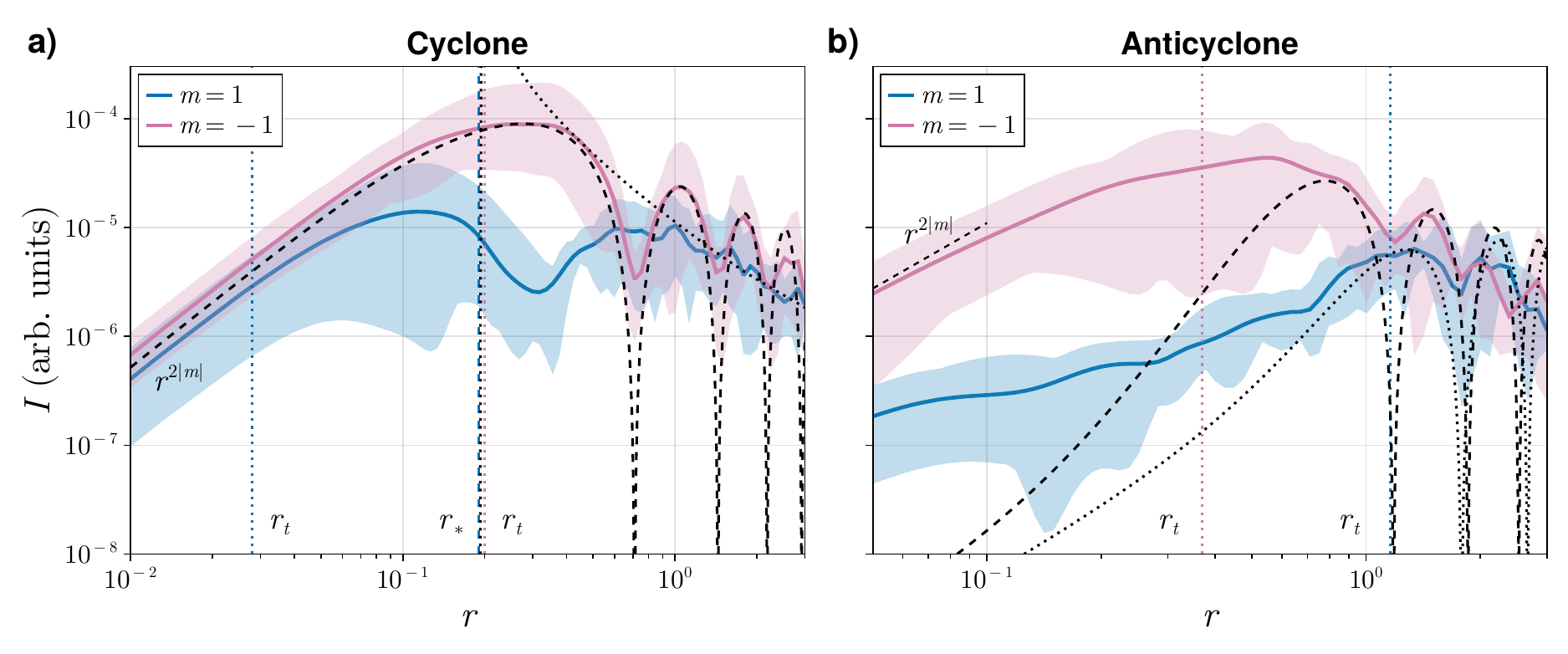}
    \caption{Radial profiles of the spectral intensity of the $m=\pm1$ inertial-wave modes for a cyclone (a) and an anticyclone (b), with $k_z=4$ and $\omega=10.6$. Solid lines show DNS data with standard-deviation shading, while dotted lines show numerical solutions of the wave equation (A6) of Ref.~\onlinecite{ivchenko2025absorption}. Vertical lines indicate the critical layer or turning point, where applicable.}
    \label{fig:10}
\end{figure*}

The results of the wave analysis are presented in Figs.~\ref{fig:9} and \ref{fig:10}. We first consider the axisymmetric mode $m=0$, shown in Fig.~\ref{fig:9}, for which the Doppler shift is absent. For a cyclone, $k_r^2(r)$ increases toward the vortex axis, corresponding to an effective potential well in the wave equation (\ref{eq:m=0}). In contrast, an anticyclone produces an effective potential elevation near the axis. This difference explains two key features of the wave field: the wave amplitude increases toward the center of the cyclone while remaining approximately constant within the anticyclone, and the oscillations become increasingly rapid (note log-scale) in the cyclone but cease near the axis of the anticyclone.

To support this qualitative analysis, we numerically solve Eq.~(A6) of Ref.~\onlinecite{ivchenko2025absorption}; the resulting $|w|^2$ is shown by the dashed lines in Fig.~\ref{fig:9}.  We consider a standing-wave solution, corresponding to a converging cylindrical wave that is reflected at the axis $r=0$ and returns as a divergent wave. For the cyclone, the analytical solution agrees with the DNS data within their standard deviation, except in the vicinity of the nodes. For the anticyclone, however, the analytical solution substantially underestimates the wave amplitude near the axis. This discrepancy arises because Eq.~(A6) does not account for the distributed forcing acting throughout the vortex. In the cyclone, this forcing has a relatively minor effect because the wave amplitude increases toward the center. In the anticyclone, by contrast, the forcing becomes important because the solution decreases toward the axis. The same forcing also prevents the wave amplitude in the DNS from vanishing completely at the nominal nodes.

The axisymmetric modes $m=0$ carry no angular momentum and thus do not contribute to maintaining the flow inside the vortices. We therefore focus here on the $m=\pm1$ modes, whose radial profiles are shown in Fig.~\ref{fig:10}. Corresponding results for $m=\pm2$ and $\pm3$ are provided in the SM~\cite{SM}. Inside the cyclone, the $m=1$ mode can develop a critical layer according to Eq.~(\ref{eq:crl}). For the chosen values of $\omega$ and $k_z$, such a layer is indeed present at $r_\ast\approx0.2$, as indicated in Fig.~\ref{fig:10}a. For $r>r_\ast$, the mode forms a converging rather than a standing wave because it is absorbed at the critical layer. Accordingly, the running-wave solution has no nodes, consistent with the DNS data. The numerical solution of the wave equation~(A6), shown by the dotted line, agrees well with the DNS data for $r>1$. In the region $r_\ast<r<1$, viscous effects become important, as the estimated thickness of the viscous sublayer is $(r/m\Sigma)(2\widetilde\Omega(2\widetilde\Omega+\Sigma)\nu k_z^2)^{1/3}\sim0.5$; see Ref.~\onlinecite{ivchenko2025absorption} for details. Beyond the critical layer, the wave-equation solution becomes very small, making the local external forcing significant. For the opposite mode $m=-1$ no critical layer is present; instead, the wave encounters a turning point at $r_t\approx0.2$. The numerical solution agrees well with the DNS data over the entire radial range.

Inside the anticyclone, the $m=-1$ mode has no critical layer for the chosen $\omega$ and $k_z$, according to Eq.~(\ref{eq:crl}), but instead encounters a turning point at $r_t\approx0.35$. The $m=1$ mode has a turning point at $r_t\approx1.0$. For both modes, the solution of the wave equation (A6) substantially underestimates the wave amplitudes at small radii due to the unaccounted forcing. To better resolve the wave behavior at large radii, we have increased the lower radial limit shown in Fig.~\ref{fig:10}b. Consequently, the small-$r$ asymptotic behavior $\propto r^{2|m|}$ is truncated in the plot, although it is observed for all curves.

The absence of a critical layer does not imply that the anticyclone cannot selectively absorb angular momentum from the waves. A subcritical mechanism, arising from the combined effects of nonlinear interaction with the mean flow and viscosity, produces distributed wave attenuation. The attenuation of a wave propagating from the vortex periphery $r=r_0$ toward its turning point is characterized by
\begin{equation}
    I_\nu 
    =
    \int\limits_{r_t}^{r_0} 
        \frac{dr}{|{\mathrm v}_{\!g}|}\, 
        \nu k^2,
\end{equation}
where ${\mathrm v}_{\!g} = \partial\widetilde\omega /\partial k_r$ is the radial group velocity and the dispersion relation is given by Eq.~(\ref{eq:dispersion}). The intensity of the reflected wave returning to the periphery is reduced by a factor $\exp(-4I_\nu)$. For $r_0=3$, we obtain attenuation factors of $0.66$ and $0.04$ for the $m=1$ and $m=-1$ modes, respectively. For comparison, the corresponding factor for the $m=-1$ mode in the cyclone is $0.73$. Thus, in the anticyclone, the wave carrying negative angular momentum ($m=-1$) is absorbed predominantly. Notably, this absorption is distributed throughout the region $r>r_t$, rather than being localized at a critical layer. In stratified fluids, a similar subcritical selective-absorption mechanism leads to quasi-biennial oscillations~\cite{semin2024quasi}.

In summary, the DNS results support the theoretical picture of inertial-wave propagation, reflection, and selective absorption by geostrophic vortices. The selection can be implemented due to presence of critical layer or subcritically due to action of viscosity. Inside the cyclones, the absolute value of the local Rossby number $\zeta_U/(2\Omega)$ is larger than inside the anticyclones. As a consequence, the resonance condition (\ref{eq:crl}) is met statistically more often inside cyclones than inside anticyclones. In anticyclones, a wider fraction of waves experiences selective absorption through distributed viscous attenuation even in the absence of critical layers. Wave with higher azimuthal harmonics or relatively low frequencies can still be absorbed at critical layers, as demonstrated in the SM~\cite{SM}. Selective absorption preferentially attenuates waves whose azimuthal wavenumbers have the appropriate sign. This asymmetry generates a nonzero off-diagonal component $\langle u^{\varphi} u^{{r}} \rangle_{Z,\Phi,T}$ of the Reynolds stress tensor, which sustains the geostrophic vortices in accordance with Eq.~(\ref{eq:Reynolds}). Although the linear wave equation captures the main features of the radial wave profiles, discrepancies near the vortex axis highlight the importance of distributed forcing in the DNS.

\section{Suppression of Geostrophic Turbulence at Rapid Rotation}

We now investigate how the degree of two-dimensionalization depends on the background rotation rate $\Omega$. As a measure of two-dimensionalization, we consider the fraction of the total kinetic energy contained in the horizontal geostrophic flow $E_{2D}/E$. At weak rotation, the Coriolis force has little influence on the flow, which therefore resembles conventional three-dimensional turbulence with a small geostrophic energy fraction. As $\Omega$ increases, rotation progressively suppresses nonlinearity in three-dimensional motions, leaving only nonlinear interaction between the three-dimensional and two-dimensional parts of the flow. As a result, this leads to increase in $E_{2D}/E$. However, there exists an optimal rotation rate for two-dimensionalization, as excessively rapid rotation leads to a decoupling between the geostrophic flow and inertial waves~\cite{gome2026waves}. 

Consistent with this picture, as shown below, the geostrophic energy fraction in our DNS initially increases with the rotation rate but decreases at sufficiently rapid rotation. This decoupling can be understood in terms of the selective wave-absorption mechanism described above. The forcing excites inertial waves with frequencies $\omega<2\Omega$ and an approximately uniform spectral density, see Fig.~\ref{fig:8}. The characteristic velocity of the geostrophic flow is bounded by the energetic estimate $U_g\lesssim\sqrt{\varepsilon/\alpha}$, where $\varepsilon$ is the energy injection rate and $\alpha$ is the linear friction coefficient. Consequently, only inertial waves satisfying $\omega\lesssim |m| k_f U_g\lesssim |m|k_f\sqrt{\varepsilon/\alpha}$ can efficiently meet the resonance condition (\ref{eq:crl}) and transfer angular momentum to the geostrophic flow. The same argument applies to subcritical selective absorption, since the wave equations for modes with opposite azimuthal wavenumbers $m$ become increasingly similar when $\omega \gg |m|k_f U_g$. Since the forcing spectrum is nearly flat in frequency, increasing $\Omega$ enlarges the total interval of excited frequencies while leaving the resonant frequency range essentially unchanged. As a result, the fraction of excited waves capable of sustaining the geostrophic vortices decreases with increasing $\Omega$, leading to a suppression of the large-scale geostrophic flow. 

Note that our situation differs fundamentally from that considered in Ref.~\onlinecite{seshasayanan2018condensates}, where the forcing is vertically invariant and therefore injects energy directly into the geostrophic modes. In that case, rapid rotation does not suppress the geostrophic flow. This distinction is also reflected in the energy flux $\Pi(k)$. In our system, the upscale energy transfer at $k<k_f$ is dominated by three-dimensional modes, see Fig.~\ref{fig:4}, whereas in Ref.~\onlinecite{seshasayanan2018condensates}, the upscale transfer in the vortex-condensate regime is carried primarily by the geostrophic modes.

\begin{figure}[b]
    \centering
    \includegraphics[width=0.9\linewidth]{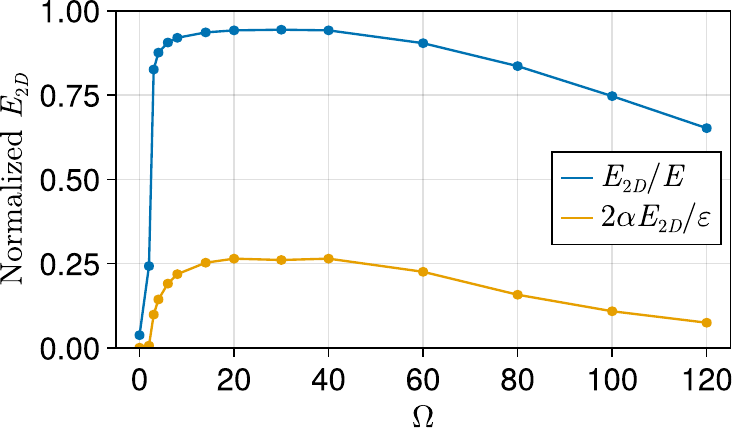}
    \caption{Normalized energies of the horizontal geostrophic flow, $E_{2D}/E$ and $2 \alpha E_{2D}/\varepsilon$, as functions of the background rotation rate $\Omega$.}
    \label{fig:11}
\end{figure}

We performed a series of DNS with the background rotation rate varied over the range $0\leq\Omega\leq120$. All simulations are initialized from a quiescent state and evolved until a statistically stationary regime is reached, as determined by the saturation of the total kinetic energy. The resulting dependencies of $2 \alpha E_{2D}/\varepsilon$ and $E_{2D}/E$ on $\Omega$ are shown in Fig.~\ref{fig:11}. In agreement with the above arguments, the geostrophic energy and its fraction in the total energy first increases with the rotation rate, reaches a maximum, and then decreases as the rotation becomes sufficiently rapid. The corresponding values of $E_{2D}$ and $E$, together with the dimensionless flow parameters $Re$, $Ro$, and $Rh$, are listed in Table~\ref{tab:1}.

\begin{table}[t]
\caption{Horizontal geostrophic energy $E_{2D}$, total system energy $E$, their ratio $E_{2D}/E$, and the corresponding dimensionless numbers
$Re=U_{\mathrm{rms}}l_f/\nu$,
$Ro=U_{\mathrm{rms}}/(2\Omega l_f)$,
and $Rh=U_{\mathrm{rms}}/(\alpha L)$
for different background rotation rates $\Omega$, where
$U_{\mathrm{rms}}=\sqrt{2E}$.}
\label{tab:1}
\centering
\begin{tabular*}{\columnwidth}{@{\extracolsep{\fill}}ccccccc@{}}
\toprule
$\Omega$ & $E_{2D}$ & $E$ & $E_{2D}/E$ & $Re$ & $Ro$ & $Rh$ \\
\midrule
0   & 0.01 & 0.21 & 0.04 & 335  & $\infty$ & 20 \\
2   & 0.07 & 0.30 & 0.24 & 406  & 0.185 & 25 \\
3   & 0.99 & 1.20 & 0.83 & 810  & 0.246 & 49 \\
4   & 1.44 & 1.65 & 0.88 & 950  & 0.217 & 58 \\
6   & 1.91 & 2.11 & 0.91 & 1075 & 0.164 & 65 \\
8   & 2.19 & 2.38 & 0.92 & 1142 & 0.130 & 70 \\
14  & 2.53 & 2.70 & 0.94 & 1216 & 0.079 & 74 \\
20  & 2.65 & 2.81 & 0.94 & 1241 & 0.057 & 76 \\
30  & 2.61 & 2.76 & 0.94 & 1230 & 0.037 & 75 \\
40  & 2.65 & 2.82 & 0.94 & 1242 & 0.028 & 76 \\
60  & 2.26 & 2.50 & 0.90 & 1170 & 0.018 & 71 \\
80  & 1.58 & 1.89 & 0.84 & 1016 & 0.012 & 62 \\
100 & 1.09 & 1.46 & 0.75 & 894  & 0.008 & 54 \\
120 & 0.75 & 1.15 & 0.65 & 795  & 0.006 & 48 \\
\bottomrule
\end{tabular*}
\end{table}

\begin{figure}[b]
    \centering
    \includegraphics[width=0.9\linewidth]{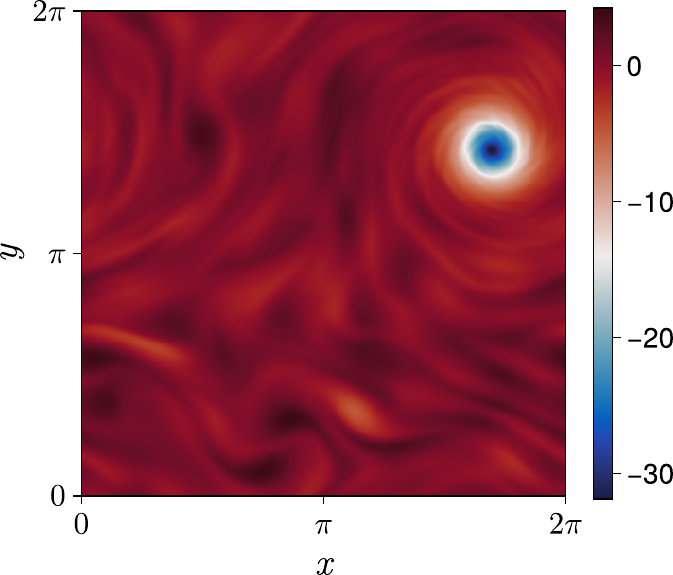}
    \caption{Snapshot of the vertical vorticity averaged along the $Z$-direction from the DNS run with $\Omega=100$.}
    \label{fig:12}
\end{figure}

Interestingly, for $\Omega \geq 60$, the system develops a single columnar vortex rather than a cyclone-anticyclone pair, see Fig.~\ref{fig:12}. This vortex is long-lived and does not decay over the duration of the simulations. Its rotation direction can be either cyclonic or anticyclonic; both cases were observed in independent simulations with identical parameters, differing only in the realization of the random forcing that drives the flow. In the limit of small local Rossby number on the vortex axis, $|\zeta_U(0)|/(2 \Omega) \ll 1$, the velocity profiles of cyclones and anticyclones become nearly identical, in agreement with the results of Ref.~\onlinecite{parfenyev2021velocity}. For the DNS run with $\Omega=100$, the local Rossby number is approximately $0.2$, and the corresponding profiles are presented in the SM~\cite{SM}.  

Finally, we note that laboratory experiments with no-slip top and bottom boundaries involve an additional mechanism that can also reduce $E_{2D}/E$ at large $\Omega$. In such systems, the inverse energy transfer is arrested by Ekman friction associated with thin boundary layers adjacent to the horizontal walls. Since the Ekman-layer thickness decreases with increasing rotation rate, the effective friction coefficient $\alpha = 2 \sqrt{\nu \Omega}/H$
increases with $\Omega$. Consequently, the upscale energy transfer is arrested at progressively smaller scales, reducing the energy accumulated in the large-scale geostrophic flow. In the present DNS, however, $\alpha$ is kept independent of $\Omega$. The suppression of geostrophic turbulence reported here therefore arises from a different mechanism: the decreasing fraction of forced inertial waves that satisfy the resonance condition and can transfer angular momentum to the geostrophic vortices.


\section{Conclusion}

We have used direct numerical simulations to examine how large-scale geostrophic vortices are maintained in rotating turbulence. By measuring the Reynolds stresses associated with inertial waves and the velocity profiles of coherent vortices, we verified that the observed large-scale flow is quantitatively consistent with the Reynolds-averaged momentum balance. We also confirm the finite-Rossby-number cyclone-anticyclone asymmetry, with cyclones exhibiting stronger and more compact cores.

Wave analysis provides evidence that vortices selectively absorb inertial waves carrying angular momentum of the same sign as their rotation. Absorption occurs either at critical layers or, when no critical layer is present, through enhanced viscous dissipation associated with the growth of the local wavenumber.

We also find that the geostrophic energy fraction depends non-monotonically on the rotation rate, in agreement with Ref.~\onlinecite{gome2026waves}. Although increasing rotation initially promotes two-dimensionalization, sufficiently rapid rotation weakens the geostrophic flow because the fraction of the forced inertial-wave spectrum capable of transferring angular momentum decreases. Thus, strong rotation does not necessarily enhance geostrophic organization; instead, it can ultimately suppress the formation and maintenance of large-scale vortices.

Of particular significance is the state observed in the regime of rapid rotation ($\Omega \geq 60$), where the flow organizes into a single, long-lived columnar geostrophic vortex rather than a cyclone-anticyclone pair. Since the horizontal boundaries are periodic, this state is not anchored to rigid walls. Moreover, the sign of its rotation is not prescribed but is selected spontaneously by the particular realization of the random forcing, with both cyclonic and anticyclonic outcomes occurring in independent runs with identical parameters. A systematic investigation of this state is left for future work.

\section*{Supplementary Material}

The supplementary material presents grid-convergence tests, examines the effects of vertical spectral treatment and Hann windowing on the energy spectra, analyzes the Reynolds-stress balance for an anticyclone, provides radial profiles of inertial-wave modes with azimuthal wavenumbers $m=\pm2$ and $m=\pm3$, and reports the velocity profiles of geostrophic vortices at $\Omega=100$.

\acknowledgments
We are grateful to N.A. Ivchenko for valuable discussions. The research presented in Sections III-VI was supported by the Basic Research Program of HSE University (project HSE-BR-2025-57), while the work reported in Sections VII-VIII was supported by the Ministry of Science and Higher Education of the Russian Federation (within the state assignment FFWR-2024-0014 of the Landau Institute for Theoretical Physics). We are also grateful to the Landau Institute for providing computing resources.

\section*{Data AVAILABILITY}
The data that support the findings of this study are available from the corresponding author upon reasonable request.

\bibliography{biblio}

\clearpage 
\onecolumngrid 
\pagestyle{empty} 
\setlength{\hoffset}{-1in}
\setlength{\voffset}{-1in}
\setlength{\topmargin}{0pt}
\setlength{\headheight}{0pt}
\setlength{\headsep}{0pt}
\setlength{\oddsidemargin}{0pt}
\noindent\includegraphics[page=1,width=\paperwidth,height=\paperheight,keepaspectratio]{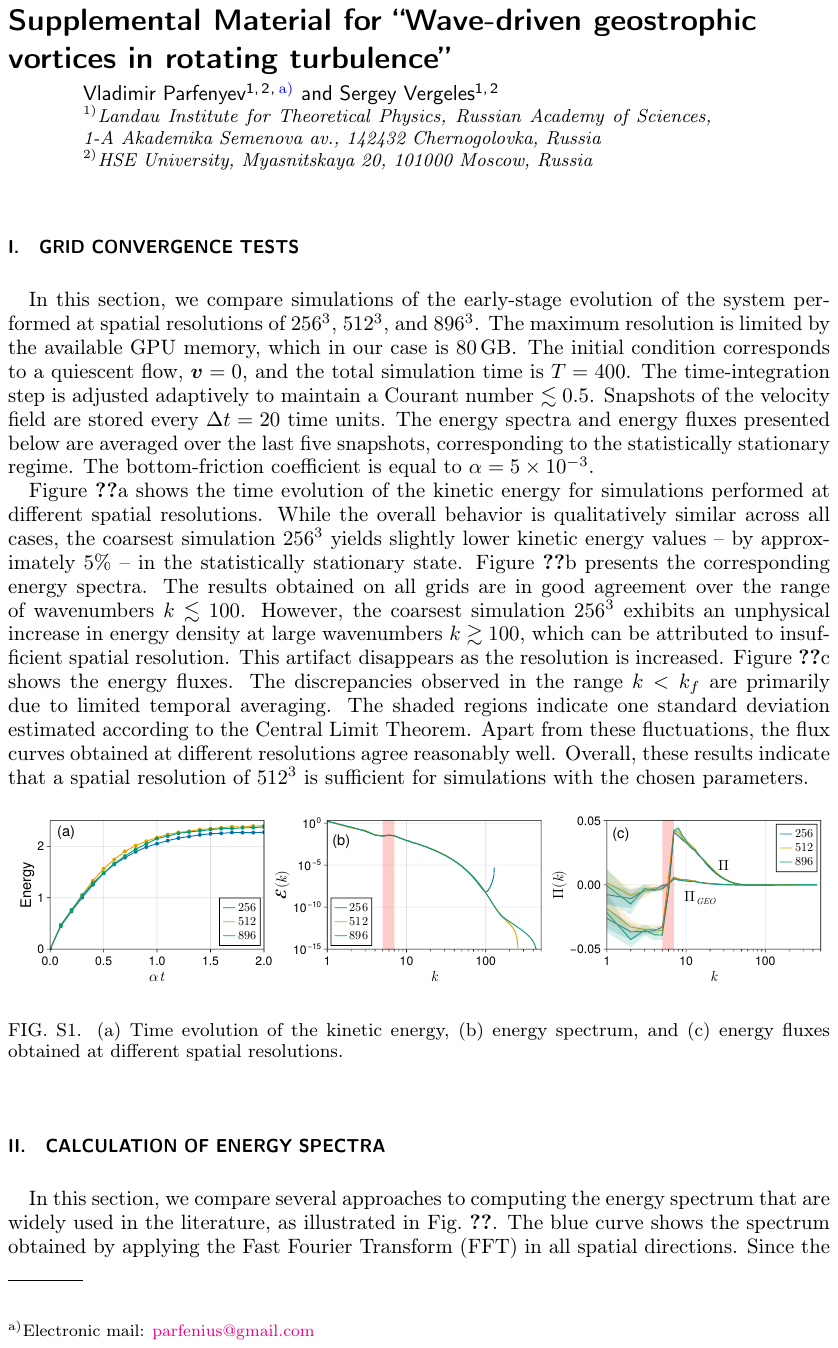}
\clearpage
\noindent\includegraphics[page=2,width=\paperwidth,height=\paperheight,keepaspectratio]{SM.pdf}
\clearpage
\noindent\includegraphics[page=3,width=\paperwidth,height=\paperheight,keepaspectratio]{SM.pdf}
\clearpage
\noindent\includegraphics[page=4,width=\paperwidth,height=\paperheight,keepaspectratio]{SM.pdf}
\clearpage

\end{document}